\documentclass[aps,prd,a4paper,twocolumn,showpacs,showkeys,preprintnumbers,amsmath,amssymb,nofootinbib,usenames,dvipsnames]{revtex4-2}
\usepackage{graphicx}
\usepackage{bm}
\usepackage{dcolumn}
\usepackage{color}

\usepackage{mathrsfs} 

\newcommand{\be}{\begin{eqnarray}}
\newcommand{\ee}{\end{eqnarray}}

\def\beq{\begin{equation}}
\def\eeq{\end{equation}}

\newcommand{\rd}{\mathrm{d}}  

\newcommand{\eq}[1]{(\ref{#1})}
\newcommand{\n}[1]{\label{#1}}

\def\al{\alpha}
\def\be{\beta}
\def\ga{\gamma}

\def\de{\delta}

\def\th{\theta}

\def\la{\lambda}

\def\rh{\rho}
\def\si{\sigma}

\def\ph{\varphi}
\def\om{\omega}

\def\Om{\Omega}

\def\na{\nabla}

\begin{document}

\title{Action principle selection of regular black holes
\vspace{0.2cm}}

\author{Breno L. Giacchini}
\email{breno@sustech.edu.cn}

\author{Tib\'{e}rio de Paula Netto}
\email{tiberio@sustech.edu.cn}

\author{Leonardo Modesto}
\email{lmodesto@sustech.edu.cn}

\affiliation{
{\small Department of Physics, Southern University of Science and Technology, Shenzhen 518055, China}
}


\begin{abstract} 
\vspace{0.1cm}
\noindent
We elaborate on the role of higher-derivative curvature invariants as a quantum selection mechanism of regular spacetimes 
in the framework of the Lorentzian path integral approach to quantum gravity. 
We show that for a large class of black hole metrics prominently regular there are higher-derivative curvature invariants which are singular. If such terms are included in the action, 
 according to the finite action principle applied to a higher-derivative gravity model, not only singular spacetimes but also some of the regular ones do not seem to contribute to the path integral.
\vspace{1cm}
\end{abstract}


\maketitle
\noindent


\section{Introduction}
\label{Sec1}

It is widely acknowledged that quantum effects would be a way of resolving the spacetime singularities present in Einstein's theory.
We hereby consider the appealing possibility that a finite action principle for gravity, consistent with good quantum properties like renormalizability and/or finiteness, may be at the core of the resolution to the singularity issue.

In the perturbative approach to quantum gravity (QG) it is common to consider higher-derivative actions, as they can result in a (super)renormalizable theory~\cite{Stelle77,AsoreyLopezShapiro}. From the classical point of view, at least in the linear approximation, higher derivatives can regularize the modified Newtonian potential~\cite{Newton-MNS,Newton-BLG,BreTib2} and the linearized curvature invariants associated with a Dirac delta source~\cite{BreTib1,BreTib2,Nos6der}. However, this is not enough to claim that singularities cannot appear in the nonlinear regime nor rule out singular solutions that couple to other energy configurations (see, for instance,~\cite{Stelle78,Stelle15-PRL-PRD,Holdom:2002xy,Holdom:2016nek,Bonanno:2019rsq}).

In the work~\cite{Lehners:2019ibe} it was remarked that higher derivatives could also have a fruitful role in the Lorentzian path integral approach to QG, in what concerns the cosmological singularity. The central assumption here is that metrics which make the action to diverge produce a destructive interference in the quantum gravity functional integral 
\beq
\n{fun}
\int D g _{\mu\nu}  
\, e^{i S [g _{\mu\nu}  ]}
.
\eeq 
Therefore, the integration over all possible geometries selects with higher weight the spacetime configurations for which the action is finite. (Earlier considerations on the finite action principle were formulated in~\citep{BarrowTipler}; see also~\cite{Barrow:2019gzc,Borissova:2020knn,Jonas:2021xkx,Chojnacki:2021ves} for a more recent discussion and further applications.) We remind that in the path integral~\eq{fun} all kinds of metrics are included, not only those which are solutions of the classical equations of motion. 

A similar reasoning was applied, in Ref.~\cite{Borissova:2020knn}, to the black hole singularities. 
The authors argued that viable gravitational actions are the ones which diverge for singular metrics.
A simple example is provided by the Schwarzschild metric, which is singular but Ricci-flat. Thus, the Einstein--Hilbert action evaluated at this solution is null, while the general fourth-derivative gravity action (see Eq.~\eq{4HD} below) diverges. Following~\cite{Borissova:2020knn}, this observation would support fourth-derivative gravity against Einstein gravity because, in the former, the singular Schwarzschild metric would be filtered out.

In the present work we elaborate more on the role played by higher-derivative invariants in the gravitational action as a selection mechanism of regular spacetimes. 
Here we go beyond the analysis of~\cite{Borissova:2020knn}, and study also the regularity of the terms {\it involving derivatives of curvature tensors}.
Our central argument is that the presence of these scalars in the action imposes further constraints on the space of regular metrics which contribute to the path integral.

The paper is organized as follows. In the next section we provide a brief review of the relevance of curvature-derivative terms already in the perturbative framework, where they are closely related to the problems of renormalizability and unitarity in QG.
In Sec.~\ref{Sec2} we show by direct calculation that some regular metrics exhibit higher-order curvature scalars that are singular. It is conjectured that for spherically symmetric spacetimes in Schwarzschild coordinates there is a relation between the occurrence of odd powers of $r$ in the metric's Taylor expansion and the existence of diverging curvature-derivative invariants; several regular black holes examples supporting the conjecture are presented.
More general results toward the conjecture are obtained in Sec.~\ref{Sec3}, where the class of invariants of type $\Box^n R$ is studied in detail.
In Sec.~\ref{Sec4} we use the results of the previous sections to evaluate the contribution of the curvature-derivative invariants to the action. It is the main result of this paper that also some regular black holes do not contribute to the Lorentzian path integral, depending on the gravitational action considered.
Finally, in Sec.~\ref{Sec5} we briefly comment on non-analytic regular metrics; our conclusions are summarized in Sec.~\ref{Sec6}.  We adopt units such that $c = \hbar = G = 1$.

\section{Derivative invariants in perturbative quantum gravity}
\label{Sec1.2}

It is instructive to begin by recalling the importance of higher-derivative terms---especially the curvature-derivative terms---in the perturbative approach to QG.
In fact, starting from the quantum version of general relativity, the power-counting analysis shows that there are diagrams with logarithmic superficial divergence
that satisfy the relation 
\beq
\n{pc-GR}
d = 2 L + 2,
\eeq  
where $d$ is the number of derivatives acting on the external graviton lines and $L$ denotes the number of loops in the diagram. Hence, from \eq{pc-GR} and the covariance of the counterterms, one can expect the presence, at the one-loop level, of the counterterms \cite{hove,dene,KTT}
\beq
\n{div-oneloop}
R_{\mu\nu\al\be}^2, 
\quad 
R_{\mu\nu}^2, 
\quad
R^2, 
\quad
\Box R
,
\eeq
($\Box = g^{\mu\nu} \na_\mu \na_\nu$ denotes the d'Alembertian) while at the two-loop level one has
\beq
\begin{split}
\n{div-2loop}
&
R_{\mu\nu\al\be}
\,R^{\al\be}_{\,\cdot\,\,\cdot\,\rho\si}
\,R^{\rho\si\mu\nu}, 
\,\,\,
R_{\mu\nu} R^{\mu\la\al\be} R^\nu_{\,\cdot\,\la\al\be}, 
\,\,
... \,\,,
\,\,
R^3,
\\
&
\hskip 0.45 cm
(\na_\la R_{\mu\nu\al\be})^2,
\,\,
... \,\,,
\,\,
R_{\mu\nu} \Box R^{\mu\nu},
\,\,\,
R \Box R,
\,\,\,
\Box^2 R,
\end{split}
\eeq 
and so on. Since the first structure in~\eqref{div-2loop} does not vanish on the mass shell~\cite{GorSag,vanVen}, the theory is not renormalizable. New types of counterterms must be introduced at every order in the loop expansion and, for each of them, it is necessary to fix a renormalization condition---which involves performing a measurement. Thus, in the usual sense, the non-renormalizability of Einstein gravity means that the theory has no predictive power.

However, general relativity is well suited for the effective theory approach~\cite{don}, because the graviton is a massless field and the next degrees of freedom due to the higher derivatives have huge masses (see, {e.g.},~\cite{Accioly:2016etf}). Indeed, performing an expansion in the energy ${\cal E}$, the higher-derivative terms are suppressed by powers of ${\cal E}/M_\text{P}$, where $M_\text{P} \sim 10^{19} \, \text{GeV}$ is the Planck mass. At high energies all the relevant operators should be considered in the gravitational action accordingly. In particular, the power counting~\eq{pc-GR} shows that at $L$-loops order it is necessary to introduce into the action terms with up to $2L + 2$ derivatives of the metric. For a recent discussion about a gauge- and parametrization-independent formulation of the renormalization group in effective quantum gravity, see~\cite{Giacchini:2020zrl}.

The renormalization of QG models depends, of course, on the choice of the initial classical action. Regarded as a fundamental (i.e., not effective) quantum field theory, the Einstein--Hilbert action $S_{\text{EH}}$  (including the cosmological~constant) augmented by fourth-derivate invariants leads to a quantum theory which is proven to be renormalizable~\cite{Stelle77}, namely
\beq
\n{4HD}
S = S_{\text{EH}} +  \int \rd^4 x \sqrt{-g} \left\{
a_1 C^2 + a_2 E + a_3 \Box R + a_4 R^2
\right\},
\eeq
where $C^2 = R_{\mu\nu\al\be}^2 - 2 R_{\mu\nu}^2 + 1/3 \, R^2$ is the square of the Weyl tensor and $E = R_{\mu\nu\al\be}^2 - 4 R_{\mu\nu}^2 + R^2$ denotes the integrand of the topological Gauss--Bonnet term in four dimensions.
The theory \eq{4HD} also represents the minimal set of terms required in the gravitational action for having a consistent quantum field theory of matter fields in curved spacetime. In fact, even if the metric is not quantized, but treated as an external classical field, the higher-derivative terms in \eq{4HD}  are generated with logarithmic divergent coefficients in vacuum diagrams (see, {e.g.}, \cite{birdav,book,Shapiro:2008sf}).

It deserves to be noted that although the Gauss--Bonnet and the $\Box R$ terms do not affect the classical dynamics of gravity, their renormalization is responsible for important quantum effects. 
For example, a significant part of the conformal (trace) anomaly is due to these structures' renormalization (see, {e.g.},~\cite{birdav,book,duff94}). 
Also, the integration of the $\Box R$-term in the conformal anomaly can generate a finite local term $\int \rd^4x \sqrt{-g} R^2$~\cite{Riegert:1984kt,
Fradkin:1983tg,Asorey:2003uf}, 
which can have fruitful applications in the construction of cosmological inflationary models~\cite{star,Starobinsky:1983zz,Fabris:1998vq,Fabris:2000gz,Pelinson:2002ef,Netto:2015cba}.
Therefore, such types of operators cannot be forgotten and must be included into the classical action entering the path integral~\eq{fun}.

One of the drawbacks of higher-derivative models are the ghost-like particles that usually exist in the spectrum if the Feynman quantization prescription is applied~\cite{Stelle77}. However,  recent developments on Lee--Wick~\cite{AsoreyLopezShapiro,ModestoShapiro16,Modesto16,AnselmiPiva1,AnselmiPiva2,AnselmiPiva3,Anselmi:2017ygm} and nonlocal theories~\cite{Krasnikov,Kuzmin,Tomboulis,Modesto12} have brought new insights on how to handle (or avoid) the ghosts and restore unitarity at any perturbative order\footnote{As the technical details regarding unitarity in these models lie beyond the scope of this work, we  refer the interested reader to the original references cited above for further considerations.}.
In both classes of models, the classical action contains \textit{more} than four derivatives of the metric.
In the former case, besides the terms in~\eq{4HD} we also include operators that are polynomial in the d'Alembertian,
\begin{eqnarray}
\hspace{-0.5 cm}  
\,\sum_{n=1}^{N} \int \rd^4 x \sqrt{-g} \,
\Big\{
\om_{n}^C \, C_{\mu\nu\al\be}\, \square^n\,C^{\mu\nu\al\be}
+ \, \om_{n}^R \, R \, \square^{n} R
\Big\},
\label{action-high}
\end{eqnarray}
where the coefficients $\om_{n}^{C,R}$ are chosen in such a way that all the
ghosts correspond to complex conjugate pairs of poles in the propagator. The unitarity of the $S$-matrix can be preserved if the complex poles are quantized {\it \`a la} Lee--Wick~\cite{AnselmiPiva2,LW1,LW2,CLOP}.

The presence of derivatives higher than four also renders this theory super-renormalizable. This can be readily seen from the power-counting formula which, instead of \eq{pc-GR}, now yields \cite{AsoreyLopezShapiro} 
\beq
\n{pcals}
d = 4 - 2 N (L-1)
\eeq 
(here $N$ is the highest power of the d'Alembertian, see Eq.~\eqref{action-high}, and we assume $\om_{N}^C, \om_{N}^R \neq 0$). Since \eq{pcals}
decreases with the number of loops, for $N = 1$ the divergences exist up to $L = 3$, while for $N \geqslant 3$ only one-loop divergences remain.  For the same reason, the theory with $N \geqslant 3$ has one-loop exact beta functions (regardless of $N$, these beta functions do not depend on the gauge and the parametrization of the quantum variables~\cite{AsoreyLopezShapiro}, as shown in~\cite{AsoreyLopezShapiro,Modesto:2017hzl,Rachwal:2021bgb}). 
We stress that for a given $N$, if there are one-loop diverging diagrams, the counterterms always have zero, two, or four derivatives of the metric. Therefore, the coefficients $\om_{n}^{C,R}$ do not get renormalized, being fixed by the choice of the initial bare classical action.

The terms quadratic in curvature in \eq{action-high} represent the minimal set of operators required for having  super-renormalizability. Since they contribute both to the propagator and the vertices, we call them {\it kinetic terms}.

In addition to the kinetic terms, there is also the possibility of including a non-minimal structure characterized by 
$O({\cal R}^3)$-terms, which, therefore, only contributes to the interaction vertices (here and in the following ${\cal R}$ denotes a generic curvature tensor). If such terms contain up to $2N+4$ derivatives of the metric tensor, the power-counting formula~\eq{pcals} does not change and the super-renormalizability of the model is not spoiled.   
For example, one can include the following operators,
\beq 
\hspace{0.6 cm}
 R^2 \Box^{N-2} R^2 \qquad \text{and} \qquad  
R_{\mu\nu} R^{\mu\nu} \Box^{N-2}  R_{\al\be} R^{\al\be}.
\eeq  
These terms can be used to kill the loop divergences by a judicious choice of their front coefficients, 
and the theory turns out to be finite at quantum level~\cite{AsoreyLopezShapiro,Modesto16,Modesto:2014lga,Modesto:2015foa}. 

It is worthwhile to mention that in principle one could also include in the action the following Gauss--Bonnet-like structures, 
\beq
\mathrm{GB}_{n}
\,=\,
R_{\mu\nu\al\be} \square^{n} R^{\mu\nu\al\be}
- 4 R_{\mu\nu} \square^{n} R^{\mu\nu}
+ R \square^{n} R,
\label{GBn}
\eeq 
which are not topological for $n \neq 0$. However, the terms in \eq{GBn} can be converted to $O({\cal R}^3)$ invariants  by means of the Bianchi identities (see, {e.g.}, \cite{AsoreyLopezShapiro,Deser:1986xr}). Therefore, they do not contribute to the flat spacetime propagator. 
Because of this, one may change the kinetic operators which are written in \eq{action-high} in the Weyl-tensor basis to another basis that uses $R_{\mu\nu\al\be}\, \square^n\,R^{\mu\nu\al\be}$ or $R_{\mu\nu}\, \square^n\,R^{\mu\nu}$ without modifying the flat-space propagator~\cite{AsoreyLopezShapiro}. The procedure is similar to what is usually done in the fourth-derivative gravity~\eq{4HD}. 

In the discussion above we emphasized the importance, in different QG models, 
of curvature invariants which can be simply polynomials in the curvature tensors or polynomials in covariant derivatives of the curvatures\footnote{Sometimes we shall refer to these scalars as ``curvature-derivative'' invariants.}.  
As already mentioned, our goal in the present work is to consider the effect of such terms in the path integral approach to QG as a mechanism for selecting regular spacetime configurations.
To this end, we here 
study the regularity of the curvature-derivative scalars evaluated at regular black hole metrics\footnote{Polynomial curvature-derivative invariants also have an interesting application in identifying the position of black hole horizons~\cite{Page:2015aia,McNutt:2017gjg,Coley:2017vxb,McNutt:2017paq}.}. 
For the sake of simplicity 
we mainly focus on invariants of the type ${\cal R} \Box^n {\cal R}$, where  we use the collective notation
\beq
{\cal R} = ( R, R_{\mu\nu}, R_{\mu\nu\al\be}, C_{\mu\nu\al\be}, \ldots ) .
\eeq


\section{Examples of singular curvature invariants for regular black holes}
\label{Sec2}

Many ``regular black hole'' metrics have been presented in the literature~\cite{Newton-MNS,Newton-BLG,BreTib2,BreTib1,Nos6der,Frolov:2016pav,Berry:2021hos,Bardeen,AyonBeato:1998ub,Dymnikova:2004zc,Nicolini:2005vd,Bronnikov:2005gm,Simpson:2018tsi,Accioly:2016qeb,Boos:2021kqe,Cano:2020ezi,Cano:2020qhy,Baake:2021jzv,Zhang14,Tseytlin:1995uq,Frolov:Exp,Frolov:Poly,Head-On,Buoninfante:2018rlq,Nicolini:2019irw,Nicolini:2012eu,Bonanno:2000ep,Hayward,Dymnikova:1992ux,DeLorenzo:2014pta,AyonBeato:1999ec,AyonBeato:1999rg,Bronnikov:2000vy,Berej:2006cc,Balart:2014cga} as exact or approximate solutions of the equations of motion in a certain classical gravity theory, as an effective treatment of quantum corrections, or motivated by other theoretical grounds. In what follows we are going to discuss some of these metrics, but before moving on it is necessary to fix the convention of what we call a ``regular'' black hole. Definitions of regular black holes can rely on the regularity of the metric components in a given  coordinate chart, on the regularity of the associated Christoffel symbols, the regularity of a set of curvature invariants, or in a limiting curvature condition (see, for instance,~\cite{Frolov:2016pav,Berry:2021hos} and references therein). In the case of the definition based on the finiteness of the curvature invariants, it is usually assumed that the curvature-squared scalars $\mathcal{R}^2$ are nonsingular. In the present work we will adopt this same definition. 
Also, we shall concentrate on static spherically symmetric metrics which may yield diverging scalars only at $r=0$, being defined by the line element
\beq \label{Met}
\rd s^2 = - A(r) \rd t^2 + B(r) \rd r^2 + r^2 \rd \Omega^2,
\eeq
where $A(r)$ and $B(r)$ are two generic functions and $\rd\Omega^2 = \rd\th^2 + \sin^2 \th \, \rd \ph^2 $ is the metric of the unit two-sphere.
Therefore,
we shall say that a metric in the form~\eqref{Met} is regular if all the invariants $R_{\mu\nu\al\be}^2$, $R_{\mu\nu}^2$, $C^2$ and $R$ are finite at $r=0$.

In this spirit, throughout this paper (except for Sec.~\ref{Sec5}) we only consider metric components which are analytic functions around $r = 0$, namely,
\beq \label{SeriesAB-0}
A(r) = \sum_{\ell=0}^{\infty} a_\ell r^\ell  , \qquad
B(r) = \sum_{\ell=0}^{\infty} b_\ell r^\ell .
\eeq
One can verify that
the conditions for regularity then read (see, {e.g.},~\cite{Frolov:2016pav} or the Sec.~\ref{Sec3} below)
\beq
\n{rego}
a_0 \neq 0, \qquad b_0 = 1,
\qquad
a_1 = b_1 = 0.
\eeq

The regularity of curvature-derivative invariants in the Newtonian limit has been studied in detail in Ref.~\cite{Nos6der}. The main result can be summarized as follows: 
{\em assuming that \eq{rego} holds and given a positive integer $n$, if the Taylor series of both functions $A(r)$ and $B(r)$ do not contain any odd power of $r$ up to (including) $r^{2n+1}$, then all the curvature-derivative invariants with at most $2n$ covariant derivatives are regular.}
For example, if the first non-zero odd coefficient occurs at $O(r^5)$, all linearized invariants of the type ${\cal R} \Box {\cal R}$ or $(\nabla_\mu {\cal R})^2$  are finite, while the first singularity might appear in the set of the operators with four covariant derivatives, such as ${\cal R} \Box^2 {\cal R}$ or $(\nabla_\mu \nabla_\nu {\cal R})^2$. This means, in particular, that if the metric components are even functions of $r$, all polynomials constructed from the curvature tensors and their derivatives are finite at $r = 0$. Moreover, the inverse implication is true, i.e.,
it is also possible to ensure the existence of some singular scalars with $2n+2$ covariant derivatives of curvatures if the first odd power in the series of the metric components is $r^{2n+3}$.

It is important to stress that the results of Ref.~\cite{Nos6der} were obtained only for the Newtonian limit, and an underlying assumption in the proof is that the curvature-derivative scalars are calculated to the lowest order in the metric perturbation. The generalization of these considerations to the full nonlinear regime (which is most relevant, since gravity is a genuinely nonlinear interaction) is considerably more involved. A thorough investigation of this problem is beyond the scope of the paper, as for our purpose it suffices to identify some diverging curvature-derivative scalars. Nonetheless, here {\em we conjecture that also in the nonlinear regime there is a relation between the occurrence of odd powers in the series expansion of the metric and the singularity of curvature invariants containing covariant derivatives}, similarly to the linearized results of~\cite{Nos6der}. In this section, therefore, we present some explicit examples supporting the conjecture---while in the next section we shall present some slightly more general results.

First, in the Table~\ref{table1} we list some regular black hole metrics that are \emph{even analytic functions}. 
According to the conjecture, since these metrics have no odd-power term in their series expansion, it is expected that all curvature-derivative invariants are regular on these spacetimes. Indeed, using \textsc{Wolfram Mathematica}~\cite{Mathematica} and the tensor algebra package \textsc{xAct}~\cite{xAct,xCoba}, for each of these metrics we evaluated a set of scalars containing up to twenty-two derivatives of a single metric component; all the invariants calculated are finite at $r=0$.  
\begin{table}[h]
\begin{tabular}{|l|c|}
\hline
Metric functions: $A(r) = 1/B(r)$ & Ref. 
\\ 
\hline
$A(r) =  1 -\frac{2Mr^2}{(r^2+\alpha^2)^{3/2}}$ & \cite{Bardeen}
\\
$A(r) =  1 -\frac{2Mr^2}{(r^2+q^2)^{3/2}} + \frac{q^2 r^2}{(r^2 + q^2)^2}$ & \cite{AyonBeato:1998ub} 
\\
$A(r) =  1-{4M\over \pi r}\left(\mbox{tan}^{-1}{r\over r_0}-
{rr_0\over r^2+r_0^2}\right)$ & \cite{Dymnikova:2004zc} 
\\
$A(r) = 1 + \frac{4M}{\sqrt{\pi} r} \, \ga ( \frac32, \frac{r^2}{4\th} )  $ & \cite{Nicolini:2005vd} 
\\
\hline
\end{tabular}
\caption{Examples of regular metrics defined by even functions for which curvature polynomials and curvature-derivative polynomial invariants are singularity-free.}
\label{table1}
\end{table}

Let us mention that other examples of spherically symmetric metrics defined by even analytic functions have been obtained in the fourth-derivative gravity (namely, the so-called $(0,0)$ family of solutions~\cite{Stelle78,Stelle15-PRL-PRD,Holdom:2016nek}) and
in the framework of nonlocal gravity models.
See, for instance, the solutions in~\cite{Tseytlin:1995uq,Zhang14,Frolov:Exp,Frolov:Poly,Head-On,Buoninfante:2018rlq,Boos:2021kqe,Nicolini:2019irw,Nicolini:2012eu}; for more general results, including logarithmic quantum corrections, the reader can consult Ref.~\cite{Nos6der}. 
We did not include these metrics in the table for the sake of simplicity, since some of them were originally obtained in coordinate choices other than~\eqref{Met}.

In what follows we present examples of regular black hole metrics for which some curvature invariants are singular, due to the occurrence of odd-power terms in their Taylor series. For these four metrics we observe that if one of the metric components has the first odd-power term at order $r^{2n+1}$, then there exist diverging scalars with $2n$ covariant derivatives, as conjectured above.


\subsection{Renormalization group improved black hole}
\label{Reuter-Bonanno}

In Ref.~\cite{Bonanno:2000ep} it was considered the improvement of the Schwarzschild metric due to the running of the Newton constant, obtained
from the flow equation for the average effective action truncated at the Einstein--Hilbert order. The resulting metric is defined by~\eqref{Met} with the functions
\beq
A (r) = \frac{1}{B(r)} = 1 - \frac{ 2 M r^2 }{r^3 + \om (r + \ga M)}
\label{ASBH}
,
\eeq
where $M$ is the mass of the black hole measured by a distant observer, and $\om$ and $\ga$ are parameters related to the running. Expanding the function $A(r)$ around $r=0$ we obtain
\beq \label{SeriesRG}
A(r) = 
1 - \frac{2}{\ga \om} \left[ r^2 - \frac{r^3}{\ga  M} + \frac{r^4}{(\ga M)^2 }  \right] + O(r^{5})
.
\eeq

It is straightforward to verify that $R$ and the quadratic-curvature scalars such as $R_{\mu\nu}^2$ and $R_{\mu\nu\al\be}^2$ are finite~\cite{Bonanno:2000ep}, as expected for a metric with a de Sitter core. For example,
\beq 
R(r) = \frac{24}{\ga \om}  - \frac{40}{\ga^2 M \om} r + \frac{60}{\ga^3 M^2 \om} r^2  + O(r^3).
\eeq
However, since the first odd power in the series $\eq{SeriesRG}$ occurs at $r^3$, as anticipated, the invariants with two derivatives of the curvature tensors may already diverge. Indeed, 
\beq \label{R-RGI}
\Box R \underset{r \to 0}{\sim} - \frac{80}{\ga^2 M \om r} ,
\eeq
while those containing higher derivatives can diverge to a higher power. For instance,
\begin{align}
&
R_{\mu\nu} \Box^2 R^{\mu\nu} \underset{r \to 0}{\sim} \frac{256}{\ga^4 M^2 \om^2 r^2}
, \label{B2}
\\
&
R_{\mu\nu} \Box^3 R^{\mu\nu} \underset{r \to 0}{\sim} \frac{256 \left(27 \ga^2 M^2+32 \ga M^2+27 \om\right)}{\ga^6 M^4 \om^3 r^2}
, \label{B3}
\\
&
R_{\mu\nu} \Box^4 R^{\mu\nu} \underset{r \to 0}{\sim} \frac{36864}{\ga^6 M^3 \om^3 r^3}
, \label{B4}
\\
&
R_{\mu\nu} \Box^5 R^{\mu\nu} \underset{r \to 0}{\sim} \frac{221184}{\ga^6 M^3 \om^3 r^5}
. \label{B5}
\end{align}
The same singularity pattern presented in Eqs.~\eqref{B2}-\eqref{B5} occurs for the invariants $R_{\mu\nu\al\be} \Box^n R^{\mu\nu\al\be}$ and $C_{\mu\nu\al\be} \Box^n C^{\mu\nu\al\be}$, while all $\Box^n R$ and $R \Box^n R$ we have evaluated diverge like $1/r$.


\subsection{Hayward spacetime}

The Hayward black hole~\cite{Hayward} is defined by the function
\beq \label{Met_Hay}
A(r) = \frac{1}{B(r)} = 1 - \frac{ 2  M  r^2}{r^3 + 2L^3},
\eeq
where $L>0$ is a parameter related to the de Sitter center. The behavior of the metric (\ref{Met_Hay}) for small $r$ differs from the (\ref{ASBH}). For (\ref{Met_Hay}) the leading odd-order term in the series expansion occurs at the order $r^5$:
\beq \label{SeriesH}
A(r) =  1 -  M \left[ \frac{r^2}{L^3} - \frac{r^5}{2L^6} + \frac{r^8}{4L^9} + O(r^{11}) \right].
\eeq
It is not difficult to verify that the associated Ricci scalar,
\beq \label{R-Hay}
R = - \frac{24 L^3  M(r^3 - 4 L^3 )}{(r^3 + 2L^3)^3}  = \frac{12 M}{L^3} + O(r^3),
\eeq
is regular, as well as the curvature-squared scalars. In Ref.~\cite{Borissova:2020knn} the calculation of some regular scalars containing up to eight derivatives of the metric was presented, and it was shown, {e.g.}, that also the six-derivative scalar $(\na_\rh R_{\mu\nu\al\be})^2$
is nonsingular for the Hayward spacetime.
Nonetheless, another scalar containing six derivatives of the metric diverges,
\beq \label{Box2R_H}
\Box^2 R \underset{r \to 0}{\sim} - \frac{504  M}{L^6 r} .
\eeq 
If one is worried that (\ref{Box2R_H}) is a total derivative, there exist other singular invariants with eight and more derivatives of the metric that diverge too. For example, 
\beq \label{RBox2R_H}
R \Box^2 R \underset{r \to 0}{\sim} - \frac{6048  M^2}{L^9 r} .
\eeq
Invariants with more derivatives can diverge even more strongly, {e.g.},
\beq
\left( \na_\al\na_\be\na_\ga\na_\de\na_\varepsilon R_{\mu\nu\rh\si} \right) ^2 \underset{r \to 0}{\sim}  \frac{304560 M^2}{L^{12} r^4}  ,
\eeq
\beq
\n{box7}
R_{\mu\nu\al\be} \Box^7 R^{\mu\nu\al\be}  \underset{r \to 0}{\sim} -\frac{6065280}{L^{12} M^3 r^3}.
\eeq


\subsection{Dymnikova spacetime}

A similar situation occurs for the Dymnikova metric~\cite{Dymnikova:1992ux}, defined by~\eqref{Met} with the function
\beq
A (r) = \frac{1}{B(r)} = 1 - \frac{2 M ( 
1 - e^{ - \frac{r^3}{L^3} } 
)}{r}
,
\eeq
where $L \equiv 2M r_0^2$ and $r_0$ is a constant related to the de Sitter core. The power series of $A(r)$ around $r=0$ reads: 
\beq \label{SeriesD}
A(r) = 
1 - 2 M \left[ \frac{r^2}{L^3} - \frac{r^5}{2 L^6 } +  \frac{r^8}{6 L^9 }
+O(r^{11}) \right] 
.
\eeq
Since the series~\eqref{SeriesH} and~\eqref{SeriesD} coincide to the order $r^6$ (up to a factor 2 associated with $M$), it follows that the small-$r$ behavior of the scalars $\Box^2 R$ and $R \Box^2 R$ for the Dymnikova spacetime are given by the Eqs.~\eqref{Box2R_H} and~\eqref{RBox2R_H} upon replacing $M \mapsto 2M$. Higher-derivative invariants have diverging behavior similar to the ones of the Hayward metric, but with different coefficients.
Therefore, although $R$
and the quadratic scalars are bounded~\cite{Dymnikova:1992ux}, there are singular scalars with six and more derivatives of the metric.


\subsection{Modified Hayward spacetimes}

Now we consider an example of metric for which $A (r) \neq 1/B(r)$.
In Ref.~\cite{DeLorenzo:2014pta} it was proposed a modification of the Hayward metric that includes the leading infrared (IR) one-loop quantum corrections to the Newtonian potential~\cite{Duff:1974ud,Donoghue:1993&94}\footnote{See also~\cite{Nos4der,Nos6der} for a recent discussion concerning the universality of this correction in higher-derivative gravity models.}
\beq
\phi (r) = - \frac{M}{r} \left( 1 + \beta \frac{1}{r^2} \right).
\eeq
The metric has the form
\beq \label{MetModHay}
A(r) = \left( 1 -  \frac{ 2 M \al \be }{r^3 \al + M \be} \right) \frac{1}{B(r)} ,
\eeq
\beq
B(r) =  \left(1 - \frac{ 2  M  r^2}{r^3 + 2L^3} \right)^{-1} .
\eeq
Bearing in mind the Hayward metric~\eqref{Met_Hay}, the extra factor in $A(r)$ has the property of encoding a non-trivial time delay between $r=0$ and an observer in the infinity because $A(0) = 1 - 2 \al \neq 1$.

Expanding $A(r)$ around $r=0$ we get:
\beq \label{SeriesHMod}
A(r) = 
1 - 2 \al - \frac{(1-2\al)M}{L^3} r^2 + \frac{2\al^2}{ M \be} r^3 + O(r^5)
.
\eeq
Differently from the Hayward metric~\eqref{Met_Hay}, now the first odd-power term occurs already at order $r^3$ [{cf.} Eq.~\eqref{SeriesH}], whereas for $B(r)$ it is still at $O(r^5)$ because this function is unchanged. By direct computation one can verify that, although $R$ and the curvature-squared scalars are bounded at $r=0$, 
\beq \label{BR-Hmod}
\Box R \underset{r \to 0}{\sim} - \frac{48 \al^2}{(1 - 2\al) \be M r} 
\eeq
diverges, and other curvature-derivative invariants have singular behavior alike those discussed in Sec.~\ref{Reuter-Bonanno}.

Another class of modified Hayward metrics with non-trivial time shift, introduced in~\cite{Frolov:2016pav}, has a similar feature. If the series of $A(r)$ contains the power $r^3$ then $\Box R$ already diverges; otherwise, it is necessary to apply two d'Alembertians to get a singular scalar, like in the case of the original Hayward metric. This happens because the series of $B(r)$ is unchanged and does contain the term $r^5$.


\section{On the existence of singular curvature invariants}
\label{Sec3}

In the previous section we provided explicit examples of regular black hole metrics with singular curvature-derivative invariants. Since the calculation of a large number of curvature scalars is a cumbersome task, here we present some more general considerations on the existence of singular curvature invariants for a given regular metric in the form~\eqref{Met}. Our main goal is to show that under certain conditions it is possible to relate the occurrence of the first odd-power in the Taylor expansion of the metric functions and the existence of diverging curvature-derivative scalars. This can also be regarded as a first step toward the conjecture formulated in the previous section. We stress that the considerations of this section represent a progression in the conclusions of Ref.~\cite{Nos6der} which were valid only for the weak gravitational field limit.

We start assuming that $A(r)$ and $B(r)$ are smooth functions that have a Taylor series representation around $r=0$,
\beq \label{SeriesAB}
A(r) = a_0 + \sum_{\ell=1}^{\infty} a_\ell r^\ell  , \qquad
B(r) = 1 + \sum_{\ell=1}^{\infty} b_\ell r^\ell .
\eeq
Moreover, let us assume that their first odd power are, respectively, $r^{2n_a+1}$ and $r^{2n_b+1}$ for some $n_a,n_b \in \mathbb{N}$. This means that $a_{2\ell +1} = 0$ for all $\ell$ smaller than $n_a$, $b_{2\ell +1} = 0 \, \forall \, \ell < n_b$, and $\,a_{2n_a + 1}, \, b_{2n_b + 1}  \neq 0$.

It is useful to introduce a notation for the property of a function $F(r)$ \emph{to have the first odd-order term in the Taylor expansion at a certain order $r^{2n+1}$}. If a function $F(r)$ has this property, we shall denote it by writing $\mathcal{P}[F] = n$. In terms of this notation, we have $\mathcal{P}[A]=n_a$ and $\mathcal{P}[B]=n_b$.

Now, consider the action of the d'Alembert operator on a smooth scalar $F(r)$,
\beq \label{BoxPh}
\Box F  = \frac{1}{B} \left[  2 \frac{F^\prime}{r} + F^{\prime\prime} - \frac{(A B^\prime - A^\prime B) F^\prime }{2 A B} \right] .
\eeq
The function $F$ could be, for example, the scalar curvature $R$, or $R_{\mu\nu}^2$.
Expanding $F(r)$ as a power series around $r=0$, 
\beq \label{SeriesPh}
F(r) = \sum_{\ell=0}^\infty f_{\ell} \, r^\ell,
\eeq
it is straightforward to verify that if $\mathcal{P}[F] \equiv n_f \geqslant 1$, then
$$\mathcal{P}[F^\prime/r] = \mathcal{P}[F^{\prime\prime}] = n_f-1,$$ whereas the first odd-order power of the last term in~\eqref{BoxPh} is at least $O(r^{2\min\lbrace n_f, n_a, n_b \rbrace + 1})$.

Hereafter, we assume $n_f <  n_a, n_b$, so that the order of the first overall odd-power term only depends on $F$. This particular choice will be justified later, when we substitute this function by the scalar curvature. In fact, under this assumption, the first odd-power term in the series of $\Box F(r)$ is
\beq
\n{G-relation}
2 (n_f+1) (2n_f+1) \, f_{2n_f+1} \, r^{2n_f-1} \neq 0.
\eeq
Therefore, $\mathcal{P}[\Box F]=n_f-1$. 
From Eq.~\eq{G-relation} one can see that if $\, n_f=0 \,$ then $\Box F$ diverges like $r^{-1}$; otherwise it is regular.
The iteration of this procedure shows that $\mathcal{P}[\Box^{n_f} F] = 0$; hence, $\Box^{n_f+1} F$ \emph{has a singularity of the type $r^{-1}$}.

Let us now turn the attention to the Ricci scalar associated with the metric in~\eqref{Met},
\beq \label{R}
R = \frac{(A^\prime)^2}{2 A^2 B} + 2 \frac{B-1}{r^2 B} + 2 \frac{A B^\prime - A^\prime B}{r A B^2} + \frac{A^\prime B^\prime - 2 A^{\prime\prime} B}{2 A B^2}.
\eeq
The requirement that $R$ is regular immediately gives\footnote{The choice $B(0) = 1$ assumed in~\eqref{SeriesAB} is also necessary for the regularity of $R$.}
\beq
\n{G-rel2}
a_0 \neq 0 \,\qquad \text{and} \qquad \, a_1 = 2 a_0 b_1.
\eeq

It is easy to verify, however, that even when \eq{G-rel2} holds for $b_1\neq 0$ the invariants $R_{\mu\nu\al\be}^2$ and $R_{\mu\nu}^2$ are singular. For example, if this is the case,
\beq
R_{\mu\nu\al\be}^2 \underset{r \to 0}{\sim} \frac{14 b_1^2}{r^2} \quad \text{and} \quad R_{\mu\nu}^2 \underset{r \to 0}{\sim} \frac{11 b_1^2}{2r^2} .
\eeq
In contrast, only if
\beq \label{Cond1} 
a_1 = b_1 = 0, 
\eeq
all the curvature scalars built only with curvature tensors (i.e., without its covariant derivatives) are finite. This is, of course, in agreement with the known results
in the literature (see, {e.g.}, \cite{Frolov:2016pav}) and with the discussion in the beginning of the Sec.~\ref{Sec2}. Therefore, since our main objective is to study
the new singularities produced when covariant derivatives act on regular quantities, in the rest of the paper we assume~\eqref{Cond1}.

A moment's reflection reveals that, in most of the cases, the first odd-order term in the Taylor expansion of $R(r)$ is at order $r^{2N-1}$, where $N\equiv\min\lbrace n_a, n_b\rbrace$. In fact, for the first term in the right-hand side of~\eqref{R}, the first odd-order power is at least $O(r^{2n_a+1})$; for the second term it is exactly at $O(r^{2n_b-1})$; for the third term it is $O(r^{2N-1})$, and for the last one it is $O(r^{2n_a-1})$. The cancellation of the $O(r^{2N-1})$-terms can take place only if the functions $A$ and $B$ are related in a very particular way.

Indeed, assuming $N = n_a < n_b$ one can explicitly evaluate the coefficient of the $O(r^{2N-1})$-term in the series of $R$, which reads,
\beq
- 2 (n_a + 1) (2n_a + 1) \frac{a_{2n_a+1}}{a_0} \neq 0,
\eeq
since $a_{2 n_a + 1} \neq 0$ by definition  [see the discussion after Eq.~\eq{SeriesAB}].
On the other hand, if $N = n_b < n_a$, the coefficient of the $O(r^{2N-1})$-term is
\beq
4 (n_b + 1) b_{2n_b+1} \neq 0.
\eeq
The remaining scenario is when $n_a = n_b = N$. In this case the first odd-power term is 
\beq \label{Coef_nA=nB}
- 2 (N+1) \left[  ( 2N + 1 ) \frac{a_{2N+1}}{a_0} - 2 b_{2N+1}  \right] r^{2N-1},
\eeq
which vanishes if and only if 
\beq \label{Relacao-ab}
a_{2N+1} =  \frac{2 a_0 b_{2N+1}}{2N + 1}.
\eeq
Solely for this particular combination of coefficients the term $r^{2N-1}$ is absent in the Taylor representation of $R(r)$. However, higher odd powers can occur depending on the other coefficients of $A(r)$ and $B(r)$. It is important to remark that the relation~\eqref{Relacao-ab} cannot be satisfied for metrics with $A(r) = 1/B(r)$ because, although $n_a = n_b$, in this case the coefficients satisfy the relation $a_{2N+1} = - a_0^2 b_{2N+1}$, which is not compatible with~\eqref{Relacao-ab}.

\begin{figure}
\begin{center}
\includegraphics[scale=0.74,angle=0]{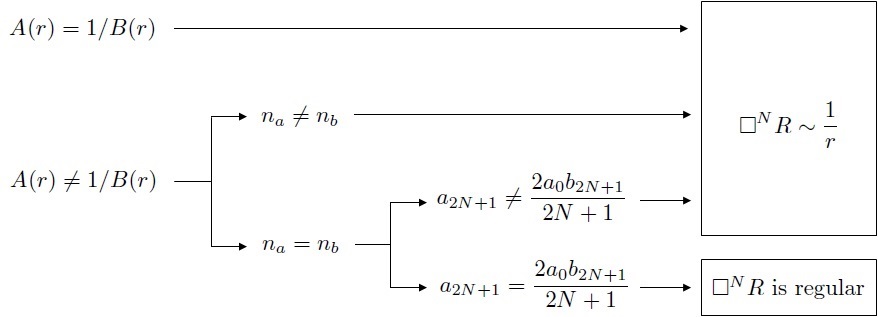}
\caption{ \sl Summary of the results on the singularity of $\Box^N R$ depending on the occurrence of the first odd power in the Taylor series of the functions $A(r)$ and $B(r)$. Following the notation in the main text, $N\equiv\min\lbrace n_a, n_b\rbrace$.}
\label{Fig1}
\end{center}
\end{figure}

Hence, unless~\eqref{Relacao-ab} is satisfied, the Ricci scalar is such that 
\beq
\mathcal{P}[R(r)] = N - 1  < n_a, n_b.
\eeq 
In particular, the hypothesis $n_f < n_a, n_b$ evoked in the considerations concerning the regularity of $\Box F$ is satisfied by the scalar $F = R$. Thus, from the previous results
it follows that $\Box R \sim r^{-1}$ for $N=1$, while if $N \geqslant 2$ we have  
$\mathcal{P}[\Box R] = N-2$. Furthermore, $\mathcal{P}[\Box^{N-1} R] = 0$ and, thus, $\Box^{N} R$ has a singularity of the type $r^{-1}$. A schematic summary of the results obtained in this section is given in the Fig.~\ref{Fig1}.

Some of the examples presented in Sec.~\ref{Sec2} can be derived from this general result. Indeed, for the metrics of the subsections A, B and C we immediately get $\mathcal{P}[R(r)] = N - 1$, inasmuch as they have $B(r)=1/A(r)$. For the Hayward and Dymnikova black holes, the first odd-order term in the series of the metric is $r^5$ ({i.e.}, $N=2$), thus $R$ and $\Box R$ are finite but $\Box^2 R$ is singular. In contrast, $N=1$ for the renormalization group improved black hole metric, so $\Box R$ already diverges. Finally, for the modified Hayward metric~\eqref{MetModHay} of subsection D we have $n_a = 1$ and $n_b=2$; whence $N=n_a=1$ and $\Box R$ is singular at $r=0$.


\section{Singularities of superficial and kinetic curvature invariants}
\label{Sec4}

In the previous section we showed that if the series expansion of the metric~\eqref{Met} around $r=0$ contains the first odd-power term at order $r^{2N+1}$, given a regular curvature invariant $F(r)$ such that $\mathcal{P}[F]=n_f < N$, the operator $\Box^{n_f+1} F$ might diverge like $r^{-1}$. This is the case for the scalar curvature if the relation~\eqref{Relacao-ab} does not hold. Moreover, if $N\geqslant 1$, all the terms of the type $R \Box^n R$ for $n < N$ are finite; while $R \Box^{N} R$ may diverge like $r^{-1} $ as $r \to 0$. 

However, from the point of view of the finite action principle, singularities of the type $r^{-1}$ and $r^{-2}$ are harmless in four-dimensional spacetimes because they do not produce a divergent action. In fact, the contribution of a given curvature invariant $F(r)$ to the gravitational action is through the term
\beq \label{ultimaSQN}
S_F (\Om_{\text{UV}},\Om_{\text{IR}}) = \int_{\Om_{\text{UV}}}^{\Om_{\text{IR}}} \rd r \sqrt{-g^{(2)}}  \,  F(r) ,
\eeq 
where, following the procedure of~\cite{Borissova:2020knn}, we introduced infrared and ultraviolet cutoffs and restrict considerations to the integral on the coordinate $r$, which is the only relevant one for static spherically symmetric metrics. Since the factor
\beq
\sqrt{-g^{(2)}} = \sqrt{A(r)B(r)}  \, r^2 = \sqrt{a_0} r^2 + O(r^4),
\eeq
which comes from the two-dimensional measure, is $O(r^2)$ for a regular metric, only if $F(r)$ is more singular than $r^{-2}$ the action term $S_F$ will diverge when we let $\Om_{\text{UV}}\to 0$.
Thus, our goal in this section is to investigate whether there are scalars of the type $\Box^n F$, or ${\cal R} \Box^n {\cal R}$, with $n>N$ that diverge with a higher inverse power of~$r$.
  
To this end, we assume that a generic curvature invariant $I(r)$ has a $r^{-1}$-type of singularity, {i.e.}, 
\beq \label{Si}
I(r) = \frac{c}{r} + F(r) ,
\eeq
where $c\neq 0$ is a constant and  $F(r)$ is an analytic function given by the power series~\eqref{SeriesPh}.
Hence, when we apply the d'Alembert operator on it we get, for $r \neq 0$ [see Eq.~\eqref{BoxPh}]:
\beq \label{BoxSi}
\begin{split}
\Box I &=  \frac{c(A B^\prime - A^\prime B)}{2 A B^2 r^2} + \Box F 
\\
&= \frac{c(a_0 b_1 - a_1)}{2 a_0 r^{2}} + O \left( 1/r \right).
\end{split}
\eeq
Therefore, only if $a_0 b_1 \neq a_1$ it happens that $\Box I \sim r^{-2}$. Otherwise, it can either be regular or diverge as $r^{-1}$, depending on the value of the coefficients $c$, $a_1$, $a_2$, $b_1$ and $b_2$, and also on $F$. 
Insomuch as $a_1 = b_1 = 0$ for regular metrics, the application of this result iteratively to the scalar $\Box^N R$---which has a singularity of the type $r^{-1}$ in the form of~\eqref{Si}---leads to the following conclusion: {\it given a regular analytic metric, 
$\Box^n R$ (for any power $n$) can only diverge like~$r^{-1}$ for $r \to 0$}.
In the sense of distributions, however,  for $n > N$ there can be singularities involving Dirac deltas and its derivatives.

The regularity of the kinetic scalars of the type $R \Box^n R$ also depends on the behavior of $R$ near $r=0$. Nevertheless, if $R$ is regular, $R \Box^n R$ cannot be more diverging than $\Box^n R$. As an important particular case, if the metric has $B=1/A$ and has a de Sitter core, {i.e.}, $a_1=b_1=0$ and $b_2 = - a_2 \neq 0$, then $R(0) = -12 a_2$ and $R \Box^n R$ always diverges as $r^{-1}$ if $\Box^n R$ diverges. This explains why we found only $1/r$-type of divergences for these invariants in the explicit evaluations considered in Sec.~\ref{Sec2}.  

However, the explicit examples of Sec.~\ref{Sec2} have also shown that the kinetic terms which are not quadratic in the Ricci scalar can diverge faster than $r^{-1}$. Consider, for instance, the invariant
\beq
K_n \equiv R_{\mu\nu\al\be} \Box^n R^{\mu\nu\al\be}.
\eeq 
For simplicity, let us also assume that $B=1/A$. If $a_3 \neq 0$, the invariant $K_1$ diverges like $r^{-1}$, but $K_2$ and $K_3$ go with $r^{-2}$, $K_4 \sim r^{-3}$ and $K_5 \sim r^{-5}$. If one sets $a_3=0$, then all these five invariants may diverge only like $r^{-1}$; nonetheless, pushing the calculations further one finds $K_6 \sim  a_4 a_5^2 r^{-2}$ and $K_7 \sim  a_4 a_5^2 r^{-4}$. If, in addition, $a_4 = 0$, then the lowest-order kinetic invariant diverging stronger than $r^{-1}$ is $K_7 \sim a_5^3 r^{-3}$; this is the case of the Hayward and Dymnikova metrics, see Eq.~\eq{box7}.
A similar situation occurs for the kinetic scalars quadratic in the Ricci or Weyl tensors [see, for instance, Eqs.~\eqref{B2}-\eqref{B5}].

{\em Therefore, we conclude that even 
for regular analytic spacetimes, 
curvature-derivative terms can yield a diverging action.} Still, as mentioned above, it seems that there exists a relation between the order of the first odd-power term in the Taylor expansion of the metric and the minimal power of the d'Alembertian in a generic kinetic scalar needed to produce a divergence stronger than~$r^{-2}$. Such a relation, though, appears to be less immediate than the one associated to the invariants $\Box^n R$ and its singularity of the type~$r^{-1}$, showed in the previous section.


\section{Comment regarding non-analytic metrics}
\label{Sec5}

So far, our considerations involved only analytic metrics in $r=0$, {i.e.}, those that can be expressed as Taylor series. However, under certain circumstances, a similar analysis can be carried out for regular metrics that are not analytic.

The most interesting scenario concerns the situation when $A(r)$ and $B(r)$ are non-analytic smooth functions. Many of these regular metrics have been obtained in the literature---see, {e.g.},~\cite{AyonBeato:1999ec,AyonBeato:1999rg,Bronnikov:2000vy,Berej:2006cc,Balart:2014cga}. The main observation here is that if $A(r)$ and $B(r)$ tend to a constant value (as $r\to 0$) faster than any polynomial, all the invariants polynomial in curvatures and their derivatives are regular. For example, consider the metric of Ref.~\cite{Bronnikov:2000vy},
\beq
A (r) = \frac{1}{B(r)} = 1 - \frac{ 2 M }{r} \left[ 1 - \tanh ( r_0/r ) \right] ,
\eeq
where $r_0$ is a positive parameter. Since
\beq
R = \frac{4 M r_0^2 \tanh(r_0/r) \text{sech}^2(r_0/r) }{r^5}
\eeq
and $\text{sech}(r_0/r)$ is exponentially suppressed by a factor $2e^{-r_0/r}$ as $r\to 0$, all the covariant derivatives of $R$ will still vanish faster than any polynomial.

Another possibility of non-analytic metrics are those which are not smooth but still admit a Taylor representation around $r=0$, with a non-polynomial remainder function. Though it is clear that the non-existence of a derivative of a certain order at $r=0$ implies that there will be singular curvature-derivative scalars, the minimal number of derivatives in such diverging scalars may be less than the order of differentiability of the functions $A(r)$ and $B(r)$. Take, for example, a metric defined by the functions $A(r)$ and $B(r)$ which are $n$-times continuously differentiable at $r=0$ (with $n \geqslant 2$).
Let us define $\mathcal{P}[A]= n_a$ and $\mathcal{P}[B]= n_b$ such that the first odd-power terms in the Taylor polynomial of $A(r)$ and $B(r)$ are respectively at order $r^{2n_a+1}$ and $r^{2n_b+1}$. Then, from the discussion of Sec.~\ref{Sec3}, if $2N \equiv 2\min\lbrace n_a,n_b\rbrace<n-1$, it can happen that
$\Box^{N} R$ diverges as $r^{-1}$. However, depending on the remainder function, higher-order invariants can have singularities which are not of the type of an inverse power of $r$, as we show in the next example.

For instance, in the context of the sixth-derivative gravity  with logarithmic quantum corrections, in Ref.~\cite{Nos6der} it was obtained a metric defined by the function
\beq \label{Series6der}
A(r) = \sum_{\ell=0}^\infty a_{\ell} r^{\ell} + \log(\mu r) \sum_{\ell=2}^\infty c_{2\ell+1} r^{2\ell+1},
\eeq
with $a_1 \equiv 0$ but $a_3, c_5 \neq 0$ ($\mu>0$ is a parameter and $B(r)$ has the same general structure).
Even though the function $A(r)$ is four-times differentiable in $r=0$, $\Box R$ already diverges like $r^{-1}$ due to the presence of the term $r^3$ in~\eqref{Series6der}, whereas $\Box^2 R \sim \log r / r$ for small $r$, as a consequence of  non-analyticity.


\section{Summary and discussion}
\label{Sec6}

In this paper we stressed that regular black hole metrics can still produce divergent curvature-derivative scalars, like $\Box^n R$,  $R_{\mu\nu\al\be} \Box^n R^{\mu\nu\al\be}$ and similar ones. This observation is based in the work~\cite{Nos6der}, where it was shown that, within the linear approximation, there is a relation between the occurrence of odd powers of $r$ in the Taylor series of a static spherically symmetric metric and the minimum  number of covariant derivatives in a given singular scalar. In particular, if such a metric exhibits an odd power of $r$, it is possible to guarantee the existence of diverging curvature-derivative invariants~\cite{Nos6der}. On the contrary, if the metric is an even analytic function of $r$, then all these invariants are bounded. In the case where the metric is nonanalytic, the finiteness of the all the curvature invariants is ensured if the metric components fall off faster than any polynomial as $r$ goes to zero.

The extension of the results of~\cite{Nos6der} beyond the linear regime is a much more involved task, but it seems that there is still a relation between odd powers of $r$ and the occurrence of singularities in derivative scalars. Indeed, in the Sec.~\ref{Sec2} we showed explicit examples of regular black hole metrics and curvature invariants corroborating this hypothesis. Some more general results were derived in the Sec.~\ref{Sec3}, focused  on the class of scalars $\Box^n R$. The main result in this regard is that if the first odd power in the Taylor series of a metric is $r^{2N+1}$ then (unless the relation~\eqref{Relacao-ab} is verified, see Fig.~\ref{Fig1}) $\Box^N R$ diverges like $r^{-1}$. The scalars of same type but with higher powers of the d'Alembertian can only diverge like $r^{-1}$. Nonetheless, invariants of the kind $R_{\mu\nu\al\be} \Box^n R^{\mu\nu\al\be}$, for example, might have stronger divergences, as discussed in Secs.~\ref{Sec2} and~\ref{Sec4}.

The aforementioned observations have a fruitful application to 
the finite-action principle in the context of the Lorentzian path integral approach to quantum gravity. In fact,
the inclusion of higher-derivative operators in a gravitational action might cause the action to diverge even when calculated on regular spacetimes. As argued in~\cite{Borissova:2020knn}, the contribution of a curvature invariant $F(r)$ evaluated at a static spherically symmetric metric in the form~\eqref{Met} is via
\beq
\label{ultima}
S_F (\Om_{\text{UV}},\Om_{\text{IR}}) = \int_{\Om_{\text{UV}}}^{\Om_{\text{IR}}} \rd r \sqrt{A(r)B(r)}  \, r^2 F(r) .
\eeq
\vspace{0.1cm}

\noindent
Since $\sqrt{AB}$ is finite for a regular metric, if $F(r)$ diverges like $r^{-n}$ with $n > 2$, then $S_F (\Om_{\text{UV}},\Om_{\text{IR}})$ will diverge when we send the UV cutoff to zero. If this is the case, following the finite action principle~\cite{Lehners:2019ibe,BarrowTipler,Barrow:2019gzc,Borissova:2020knn},  the corresponding spacetime configuration is going to be filtered out of the functional integration over all geometries.

Therefore, depending on the form of the gravitational action, it might happen that some regular spacetime configurations do not contribute to path integral.

In view of the results presented in Sec.~\ref{Sec3}, if the fourth-derivative gravity is taken as the fundamental QG theory, then for any regular black hole metric~\eqref{Met} the divergence that may exist at $r=0$ in the term $\Box R$ (which is required by renormalization) does not yield an infinite action. This happens because for these metrics such term can only diverge like $1/r$; moreover, this is the sole scalar with at most four derivatives that can diverge on a regular metric.

However, as pointed out in Ref.~\cite{Borissova:2020knn}, the requirement of finiteness may be a way of distinguishing viable gravitational actions in their capability of selecting regular geometries. In this spirit, a theory containing  higher-derivative structures like those in~\eq{action-high} with large $N$ is favored in comparison to the simple fourth-derivative gravity~\eq{4HD}. In fact, the space of regular metrics associated with a finite action is more constrained if the number of derivatives in the action is larger.


\acknowledgments
This work was supported by the Basic Research Program of the Science, Technology and Innovation Commission of Shenzhen Municipality (grant no. JCYJ20180302174206969). The authors thank the anonymous referee for the useful remarks for the improvement of the text.





\begin{thebibliography}{99}

\bibitem{Stelle77}  K.~S.~Stelle,
{\it Renormalization of Higher Derivative Quantum Gravity},
Phys.\ Rev.\ D {\bf 16}, 953 (1977).

\bibitem{AsoreyLopezShapiro} 
M.~Asorey, J.~L.~L\'opez and I.~L.~Shapiro,
{\it Some remarks on high derivative quantum gravity},
Int.\ J.\ Mod.\ Phys.\ A {\bf 12}, 5711 (1997), arXiv:hep-th/9610006.

\bibitem{Newton-MNS} 
L.~Modesto, T.~de~Paula~Netto and I.~L.~Shapiro,
{\it On Newtonian singularities in higher derivative gravity models},
J. High Energy Phys. {\bf 1504}, 098 (2015), 
arXiv:1412.0740.

\bibitem{Newton-BLG} 
B.~L.~Giacchini,
{\it On the cancellation of Newtonian singularities in higher-derivative gravity},
Phys. Lett. B {\bf 766}, 306 (2017), 
arXiv:1609.05432.

\bibitem{BreTib2} 
B.~L.~Giacchini and T.~de Paula Netto,
{\it Effective delta sources and regularity in higher-derivative and ghost--free gravity},
J. Cosmol. Astropart. Phys.  {\bf 1907}, 013 (2019),
arXiv:1809.05907.

\bibitem{BreTib1} 
B.~L.~Giacchini and T.~de Paula Netto,
{\it Weak-field limit and regular solutions in polynomial higher-derivative gravities},
Eur.\ Phys.\ J.\ C {\bf 79}, 217 (2019),
arXiv:1806.05664.

\bibitem{Nos6der}
N.~Burzill\`a, B.~L.~Giacchini, T.~de Paula Netto and L.~Modesto,
{\it Higher-order regularity in local and nonlocal quantum gravity},
Eur. Phys. J. C \textbf{81}, 462 (2021),
arXiv:2012.11829.

\bibitem{Stelle78}  
K.~S.~Stelle,
{\it Classical Gravity with Higher Derivatives},
Gen.\ Rel.\ Grav.\  {\bf 9}, 353 (1978).

\bibitem{Stelle15-PRL-PRD}
H.~L\"u, A.~Perkins, C.~N.~Pope and K.~S.~Stelle,
{\it Black Holes in Higher-Derivative Gravity},
Phys.\ Rev.\ Lett.\  {\bf 114}, 171601 (2015), arXiv:1502.01028;
{\it Spherically Symmetric Solutions in Higher-Derivative Gravity},
Phys.\ Rev.\ D {\bf 92}, 124019 (2015), arXiv:1508.00010.

\bibitem{Holdom:2016nek}
B.~Holdom and J.~Ren,
{\it Not quite a black hole},
Phys. Rev. D \textbf{95}, 084034 (2017),
arXiv:1612.04889.

\bibitem{Bonanno:2019rsq}
A.~Bonanno and S.~Silveravalle,
{\it Characterizing black hole metrics in quadratic gravity},
Phys. Rev. D \textbf{99}, 101501 (2019),
arXiv:1903.08759.

\bibitem{Holdom:2002xy}
B.~Holdom,
{\it On the fate of singularities and horizons in higher derivative gravity},
Phys. Rev. D \textbf{66}, 084010 (2002),
arXiv:hep-th/0206219.

\bibitem{Lehners:2019ibe}
J.~L.~Lehners and K.~S.~Stelle,
{\it Safe beginning for the universe?},
Phys. Rev. D \textbf{100}, 083540 (2019),
arXiv:1909.01169.

\bibitem{BarrowTipler}
J.~D.~Barrow and F.~J.~Tipler,
{\it Action principles in nature},
Nature \textbf{331}, 31 (1988).

\bibitem{Barrow:2019gzc}
J.~D.~Barrow,
{\it Finite action principle revisited},
Phys. Rev. D \textbf{101}, 023527 (2020),
arXiv:1912.12926.

\bibitem{Borissova:2020knn}
J.~N.~Borissova and A.~Eichhorn,
{\it Towards black-hole singularity-resolution in the Lorentzian gravitational path integral},
Universe \textbf{7}, 48 (2021),
arXiv:2012.08570.

\bibitem{Jonas:2021xkx}
C.~Jonas, J.~L.~Lehners and J.~Quintin,
{\it Cosmological consequences of a principle of finite amplitudes},
Phys. Rev. D \textbf{103}, 103525 (2021),
arXiv:2102.05550.

\bibitem{Chojnacki:2021ves}
J.~Chojnacki and J.~H.~Kwapisz,
{\it Finite Action Principle and Horava-Lifshitz Gravity: early universe, black holes and wormholes},
arXiv:2102.13556.


\bibitem{hove}
G.~'t Hooft and M.~Veltman,
{\it One loop divergencies in the theory of gravitation},
Ann. Inst. H. Poincar\'e Phys. Theor. A \textbf{20}, 69 (1974).

\bibitem{dene} S. Deser and P. van Nieuwenhuisen,
{ \it One-loop divergences of quantized Einstein-Maxwell fields},
Phys. Rev. D {\bf10}, 401 (1974). 

\bibitem{KTT} R. E. Kallosh, O. V. Tarasov and I. V. Tyutin,
{\it One-loop finiteness of quantum gravity off mass shell},
Nucl. Phys. B {\bf 137}, 145  (1978). 

\bibitem{GorSag} M.~H.~Goroff and A.~Sagnotti,
{\it Quantum gravity at two loops,}
Phys. Lett. {\bf B160}, 81 (1985);
{\it The ultraviolet behavior of Einstein gravity,}
Nucl. Phys. {\bf B266}, 709 (1986).

\bibitem{vanVen} A.~E.~M.~van de Ven,
{\it Two loop quantum gravity,}
Nucl. Phys. \textbf{B378}, 309 (1992). 

\bibitem{don} 
J.~F.~Donoghue,
{\it General relativity as an effective field theory:
The leading quantum corrections},
Phys. Rev. D {\bf 50}, 3874 (1994), 
arXiv:gr-qc/9405057.

\bibitem{Accioly:2016etf}
A.~Accioly, B.~L.~Giacchini and I.~L.~Shapiro,
{\it On the gravitational seesaw in higher-derivative gravity},
Eur. Phys. J. C \textbf{77}, 540 (2017),
arXiv:1604.07348.

\bibitem{Giacchini:2020zrl}
B.~L.~Giacchini, T.~de Paula Netto and I.~L.~Shapiro,
{\it On the Vilkovisky-DeWitt approach and renormalization group in effective quantum gravity},
J. High Energy Phys. \textbf{2020}, 011 (2020),
arXiv:2009.04122.

\bibitem{birdav} N. D. Birrell and P. C. W. Davies,
{\it Quantum fields in curved space}
(Cambridge University Press, Cambridge, England, 1982).

\bibitem{book}
I. L. Buchbinder, S. D. Odintsov and I. L. Shapiro, {\it
Effective Action in Quantum Gravity} (IOPP, Bristol, 1992).

\bibitem{Shapiro:2008sf}
I.~L.~Shapiro,
{\it Effective Action of Vacuum: Semiclassical Approach},
Class. Quant. Grav. \textbf{25}, 103001 (2008),
arXiv:0801.0216.

\bibitem{duff94}
M.~J.~Duff,
{\it Twenty years of the Weyl anomaly},
Class. Quant. Grav. \textbf{11}, 1387 (1994),
arXiv:hep-th/9308075.

\bibitem{Riegert:1984kt}
R.~J.~Riegert,
{\it A Nonlocal Action for the Trace Anomaly},
Phys. Lett. B \textbf{134}, 56 (1984).

\bibitem{Fradkin:1983tg}
E.~S.~Fradkin and A.~A.~Tseytlin,
{\it Conformal Anomaly in Weyl Theory and Anomaly Free Superconformal Theories},
Phys. Lett. B \textbf{134}, 187 (1984). 

\bibitem{Asorey:2003uf}
M.~Asorey, E.~V.~Gorbar and I.~L.~Shapiro,
{\it Universality and ambiguities of the conformal anomaly},
Class. Quant. Grav. \textbf{21} 163 (2004), 
arXiv:hep-th/0307187.

\bibitem{star}
A.~A.~Starobinsky, 
{\it A new type of isotropic cosmological models without singularity},
Phys. Lett. B {\bf 91}, 99 (1980).

\bibitem{Starobinsky:1983zz}
A.~A.~Starobinsky,
{\it The Perturbation Spectrum Evolving from a Nonsingular Initially De-Sitter Cosmology and the Microwave Background Anisotropy},
Sov. Astron. Lett. \textbf{9}, 302 (1983).

\bibitem{Fabris:1998vq}
J.~C.~Fabris, A.~M.~Pelinson and I.~L.~Shapiro,
{\it Anomaly induced effective action for gravity and inflation},
Grav. Cosmol. \textbf{6}, 59 (2000), 
arXiv:gr-qc/9810032.

\bibitem{Fabris:2000gz}
J.~C.~Fabris, A.~M.~Pelinson and I.~L.~Shapiro,
{\it On the gravitational waves on the background of anomaly-induced inflation},
Nucl. Phys. B \textbf{597}, 539 (2001), 
arXiv:hep-th/0009197.

\bibitem{Pelinson:2002ef}
A.~M.~Pelinson, I.~L.~Shapiro and F.~I.~Takakura,
{\it On the stability of the anomaly induced inflation},
Nucl. Phys. B \textbf{648}, 417 (2003),
arXiv:hep-ph/0208184.

\bibitem{Netto:2015cba}
T.~de Paula Netto, A.~M.~Pelinson, I.~L.~Shapiro and A.~A.~Starobinsky,
{\it From stable to unstable anomaly-induced inflation},
Eur. Phys. J. C \textbf{76}, 544 (2016),
arXiv:1509.08882.


\bibitem{ModestoShapiro16} 
L.~Modesto and I.~L.~Shapiro,
{\it Superrenormalizable quantum gravity with complex ghosts},
Phys.\ Lett.\ B {\bf 755}, 279 (2016), arXiv:1512.07600.

\bibitem{Modesto16}  
L.~Modesto,
{\it Super-renormalizable or finite Lee-Wick quantum gravity},
Nucl.\ Phys.\ B {\bf 909}, 584 (2016), arXiv:1602.02421.

\bibitem{AnselmiPiva1}
D.~Anselmi and M.~Piva,
{\it A new formulation of Lee-Wick quantum field theory},
J. High Energy Phys. \textbf{06}, 066 (2017),
arXiv:1703.04584.

\bibitem{AnselmiPiva2}
D.~Anselmi and M.~Piva,
{\it Perturbative unitarity of Lee-Wick quantum field theory},
Phys. Rev. D \textbf{96}, 045009 (2017),
arXiv:1703.05563.

\bibitem{AnselmiPiva3}
D.~Anselmi,
{\it Fakeons and Lee-Wick Models},
J. High Energy Phys. \textbf{02}, 141 (2018),
arXiv:1801.00915.

\bibitem{Anselmi:2017ygm}
D.~Anselmi,
{\it On the quantum field theory of the gravitational interactions},
J. High Energy Phys.  \textbf{06}, 086 (2017),
arXiv:1704.07728.

\bibitem{Krasnikov} 
  N.~V.~Krasnikov,
  {\it Nonlocal Gauge Theories},
  Theor.\ Math.\ Phys.\  {\bf 73}, 1184 (1987)
  [Teor.\ Mat.\ Fiz.\  {\bf 73}, 235 (1987)].

\bibitem{Kuzmin} 
  Yu.~V.~Kuz'min,
  {\it Finite nonlocal gravity},
  Sov.\ J.\ Nucl.\ Phys.\  {\bf 50}, 1011 (1989)
  [Yad.\ Fiz.\  {\bf 50}, 1630 (1989)].

\bibitem{Tomboulis} 
E.~T.~Tomboulis,
{\it Superrenormalizable gauge and gravitational theories},
arXiv:hep-th/9702146.

\bibitem{Modesto12}  
L.~Modesto,
{\it Super-renormalizable Quantum Gravity},
Phys.\ Rev.\ D {\bf 86}, 044005 (2012), arXiv:1107.2403.

\bibitem{LW1}
T.~D.~Lee and G.~C.~Wick,
{\it Negative metric and the unitarity of the S Matrix},
Nucl. Phys. B \textbf{9}, 209 (1969).

\bibitem{LW2}
T.~D.~Lee and G.~C.~Wick,
{\it Finite Theory of Quantum Electrodynamics},
Phys. Rev. D \textbf{2}, 1033 (1970).

\bibitem{CLOP}
R.~E.~Cutkosky, P.~V.~Landshoff, D.~I.~Olive and J.~C.~Polkinghorne,
{\it A non-analytic S matrix},
Nucl. Phys. B \textbf{12}, 281 (1969).

\bibitem{Modesto:2017hzl}
L.~Modesto, L.~Rachwa\l{} and I.~L.~Shapiro,
{\it Renormalization group in super-renormalizable quantum gravity},
Eur. Phys. J. C \textbf{78}, 555 (2018),
arXiv:1704.03988.

\bibitem{Rachwal:2021bgb}
L.~Rachwa\l{}, L.~Modesto, A.~Pinzul and I.~L.~Shapiro,
{\it Renormalization Group in Six-derivative Quantum Gravity},
arXiv:2104.13980.

\bibitem{Modesto:2014lga}
L.~Modesto and L.~Rachwa\l{},
{\it Super-renormalizable and finite gravitational theories},
Nucl. Phys. B \textbf{889} 228 (2014), 
arXiv:1407.8036.

\bibitem{Modesto:2015foa}
L.~Modesto, M.~Piva and L.~Rachwa\l{},
{\it Finite quantum gauge theories},
Phys. Rev. D \textbf{94}, 025021 (2016),
arXiv:1506.06227.

\bibitem{Deser:1986xr}
S.~Deser and A.~N.~Redlich,
{\it String Induced Gravity and Ghost Freedom},
Phys. Lett. B \textbf{176}, 350 (1986).

\bibitem{Page:2015aia}
D.~N.~Page and A.~A.~Shoom,
{\it Local Invariants Vanishing on Stationary Horizons: A Diagnostic for Locating Black Holes},
Phys. Rev. Lett. \textbf{114}, 141102 (2015),
arXiv:1501.03510.

\bibitem{McNutt:2017gjg}
D.~D.~McNutt and D.~N.~Page,
{\it Scalar Polynomial Curvature Invariant Vanishing on the Event Horizon of Any 
Black Hole Metric Conformal to a Static Spherical Metric},
Phys. Rev. D \textbf{95}, 084044 (2017),
arXiv:1704.02461.

\bibitem{Coley:2017vxb}
A.~Coley and D.~McNutt,
{\it Identification of black hole horizons using scalar curvature invariants},
Class. Quant. Grav. \textbf{35}, 025013 (2018),
arXiv:1710.08773.

\bibitem{McNutt:2017paq}
D.~D.~McNutt, M.~A.~H.~MacCallum, D.~Gregoris, A.~Forget, A.~A.~Coley, P.~C.~Chavy-Waddy and D.~Brooks,
{\it Cartan Invariants and Event Horizon Detection, Extended Version},
Gen. Rel. Grav. \textbf{50}, 37 (2018),
arXiv:1709.03362.


\bibitem{Frolov:2016pav}
V.~P.~Frolov,
{\it Notes on nonsingular models of black holes},
Phys. Rev. D \textbf{94}, 104056 (2016),
arXiv:1609.01758.

\bibitem{Berry:2021hos}
T.~Berry, A.~Simpson and M.~Visser,
{\it General class of ``quantum deformed'' regular black holes},
Universe \textbf{7}, 165 (2021),
arXiv:2102.02471.


\bibitem{Bardeen}
J.~M.~Bardeen,
{\it Non-singular general-relativistic gravitational collapse}, 
in Proceedings  of   International  Conference GR5, (Tbilisi University Press, Tbilisi, USSR, 1968), p. 174.

\bibitem{AyonBeato:1998ub}
E.~Ay\'on-Beato and A.~Garc\'ia,
{\it Regular black hole in general relativity coupled to nonlinear electrodynamics,}
Phys. Rev. Lett. \textbf{80}, 5056 (1998),
arXiv:gr-qc/9911046.

\bibitem{Dymnikova:2004zc}
I.~Dymnikova,
{\it Regular electrically charged vacuum structures with de Sitter centre in nonlinear electrodynamics coupled to general relativity},
Class. Quant. Grav. \textbf{21}, 4417 (2004),
arXiv:gr-qc/0407072.

\bibitem{Nicolini:2005vd}
P.~Nicolini, A.~Smailagic and E.~Spallucci,
{\it Noncommutative geometry inspired Schwarzschild black hole},
Phys. Lett. B \textbf{632}, 547 (2006),
arXiv:gr-qc/0510112,
%
P.~Nicolini,
{\it Noncommutative Black Holes, The Final Appeal To Quantum Gravity: A Review}, 
Int. J. Mod. Phys. A \textbf{24}, 1229 (2009),
arXiv:0807.1939.

\bibitem{Bronnikov:2005gm}
K.~A.~Bronnikov and J.~C.~Fabris,
{\it Regular phantom black holes},
Phys. Rev. Lett. \textbf{96}, 251101 (2006),
arXiv:gr-qc/0511109.

\bibitem{Tseytlin:1995uq}
A.~A.~Tseytlin,
{\it On singularities of spherically symmetric backgrounds in string theory},
Phys. Lett. B \textbf{363}, 223 (1995),
arXiv:hep-th/9509050.

\bibitem{Nicolini:2012eu}
P.~Nicolini,
{\it Nonlocal and generalized uncertainty principle black holes},
arXiv:1202.2102.

\bibitem{Zhang14}
Y.~Zhang, Y.~Zhu, L.~Modesto and C.~Bambi,
{\it Can static regular black holes form from gravitational collapse?},
Eur.\ Phys.\ J.\ C {\bf 75}, 96 (2015),
arXiv:1404.4770.

\bibitem{Frolov:Exp} 
V.~P.~Frolov, A.~Zelnikov and T.~de Paula Netto,
{\it Spherical collapse of small masses in the ghost-free gravity},
J. High Energy Phys. {\bf 1506}, 107 (2015), arXiv:1504.00412.

\bibitem{Frolov:Poly}
V.~P.~Frolov,
{\it Mass-gap for black hole formation in higher derivative and ghost free gravity},
Phys.\ Rev.\ Lett.\  {\bf 115}, 051102 (2015), arXiv:1505.00492.

\bibitem{Head-On}
V.~P.~Frolov and A.~Zelnikov,
{\it Head-on collision of ultrarelativistic particles in ghost-free theories of gravity},
Phys. Rev. D {\bf 93}, 064048 (2016), arXiv:1509.03336.

\bibitem{Buoninfante:2018rlq}
L.~Buoninfante, A.~S.~Koshelev, G.~Lambiase, J.~Marto and A.~Mazumdar,
{\it Conformally-flat, non-singular static metric in infinite derivative gravity},
J. Cosmol. Astropart. Phys. \textbf{06}, 014 (2018),
arXiv:1804.08195.

\bibitem{Nicolini:2019irw}
P.~Nicolini, E.~Spallucci and M.~F.~Wondrak,
{\it Quantum Corrected Black Holes from String T-Duality},
Phys. Lett. B \textbf{797}, 134888 (2019),
arXiv:1902.11242.

\bibitem{Boos:2021kqe}
J.~Boos,
{\it Non-singular ``Gauss'' black hole from non-locality: a simple model with a de Sitter core, mass gap, and no inner horizon}, 
arXiv:2104.00555.



\bibitem{Bonanno:2000ep}
A.~Bonanno and M.~Reuter,
{\it Renormalization group improved black hole space-times},
Phys. Rev. D \textbf{62}, 043008 (2000),
arXiv:hep-th/0002196.

\bibitem{Hayward}
S.~A.~Hayward,
{\it Formation and evaporation of regular black holes},
Phys. Rev. Lett. \textbf{96}, 031103 (2006),
arXiv:gr-qc/0506126.

\bibitem{Dymnikova:1992ux}
I.~Dymnikova,
{\it Vacuum nonsingular black hole},
Gen. Rel. Grav. \textbf{24}, 235 (1992).

\bibitem{DeLorenzo:2014pta}
T.~De Lorenzo, C.~Pacilio, C.~Rovelli and S.~Speziale,
{\it On the Effective Metric of a Planck Star},
Gen. Rel. Grav. \textbf{47}, 41 (2015),
arXiv:1412.6015.

\bibitem{AyonBeato:1999ec}
E.~Ay\'on-Beato and A.~Garc\'ia,
{\it Nonsingular charged black hole solution for nonlinear source},
Gen. Rel. Grav. \textbf{31}, 629 (1999),
arXiv:gr-qc/9911084.

\bibitem{AyonBeato:1999rg}
E.~Ay\'on-Beato and A.~Garc\'ia,
{\it New regular black hole solution from nonlinear electrodynamics},
Phys. Lett. B \textbf{464}, 25 (1999),
arXiv:hep-th/9911174.

\bibitem{Bronnikov:2000vy}
K.~A.~Bronnikov,
{\it Regular magnetic black holes and monopoles from nonlinear electrodynamics},
Phys. Rev. D \textbf{63}, 044005 (2001),
arXiv:gr-qc/0006014.

\bibitem{Berej:2006cc}
W.~Berej, J.~Matyjasek, D.~Tryniecki and M.~Woronowicz,
{\it Regular black holes in quadratic gravity},
Gen. Rel. Grav. \textbf{38}, 885 (2006),
arXiv:hep-th/0606185.

\bibitem{Balart:2014cga}
L.~Balart and E.~C.~Vagenas,
{\it Regular black holes with a nonlinear electrodynamics source},
Phys. Rev. D \textbf{90}, 124045 (2014),
arXiv:1408.0306.

\bibitem{Accioly:2016qeb}
A.~Accioly, B.~L.~Giacchini and I.~L.~Shapiro,
{\it Low-energy effects in a higher-derivative gravity model with real and complex massive poles},
Phys. Rev. D \textbf{96}, 104004 (2017),
arXiv:1610.05260.

\bibitem{Simpson:2018tsi}
A.~Simpson and M.~Visser,
{\it Black-bounce to traversable wormhole},
J. Cosmol. Astropart. Phys. \textbf{02}, 042 (2019),
arXiv:1812.07114.

\bibitem{Cano:2020ezi}
P.~A.~Cano and \'A.~Murcia,
{\it Resolution of Reissner-Nordstr\"om singularities by higher-derivative corrections},
Class. Quant. Grav. \textbf{38}, 075014 (2021),
arXiv:2006.15149.

\bibitem{Cano:2020qhy}
P.~A.~Cano and \'A.~Murcia,
{\it Electromagnetic Quasitopological Gravities},
J. High Energy Phys. \textbf{10}, 125 (2020),
arXiv:2007.04331.

\bibitem{Baake:2021jzv}
O.~Baake, C.~Charmousis, M.~Hassaine and M.~San Juan,
{\it Regular black holes and gravitational particle-like solutions in generic DHOST theories},
J. Cosmol. Astropart. Phys. \textbf{06}, 021 (2021),
arXiv:2104.08221.




\bibitem{Mathematica}
Wolfram Research, Inc., {\it Mathematica}, Version 8,
\texttt{https://www.wolfram.com/mathematica/}.

\bibitem{xAct}
J. M. Mart\'in-Garc\'ia, 
{\it xAct: Efficient tensor computer algebra for the Wolfram Language}, \texttt{http://xact.es}, accessed: 2021-01-27.

\bibitem{xCoba}
D. Yllanes and J. M. Mart\'in-Garc\'ia, 
{\it xCoba: General component tensor computer algebra},
\texttt{http://xact.es/xCoba/}, accessed: 2021-01-27.



\bibitem{Duff:1974ud}
M.~J.~Duff,
{\it Quantum corrections to the Schwarzschild solution},
Phys. Rev. D \textbf{9}, 1837 (1974).

\bibitem{Donoghue:1993&94}
J.~F.~Donoghue,
{\it Leading quantum correction to the Newtonian potential},
Phys. Rev. Lett. \textbf{72}, 2996 (1994),
arXiv:gr-qc/9310024;
{\it General relativity as an effective field theory: The leading quantum corrections},
Phys. Rev. D \textbf{50}, 3874 (1994),
arXiv:gr-qc/9405057.

\bibitem{Nos4der}
N.~Burzill\`a, B.~L.~Giacchini, T.~de Paula Netto and L.~Modesto,
{\it Newtonian potential in higher-derivative quantum gravity},
Phys. Rev. D \textbf{103}, 064080 (2021),
arXiv:2012.06254.



\end{thebibliography}
\end{document}